\documentstyle[12pt]{article}

\newcommand{\bY}{\mbox{\boldmath $Y$}}
\newcommand{\btheta}{\mbox{\boldmath $\theta$}}
\newcommand{\bphi}{\mbox{\boldmath $\phi$}}
\def\beq{\begin{equation}}
\def\eeq{\end{equation}}

\begin{document}

\bibstyle{apalike}
\baselineskip = 0.8cm

\title      { Testing the Markov condition \\ in ion channel recordings}
\author {       J. Timmer$^{\ast}$  \\
                S. Klein \\
                $ \quad $ \\
                Freiburger Zentrum f\"ur Datenanalyse und Modellbildung \\
                Albertstr. 26 -- 28 \\
                D - 79104 Freiburg \\
                Germany}
%
\maketitle

PACS numbers: 2.50.Ga, 87.10.+e
\vskip 2cm

$^{\ast}$
Corresponding author:\\
 J. Timmer \\
Freiburger Zentrum f\"ur Datenanalyse und Modellbildung \\
Albertstr. 26 -- 28 \\
79104 Freiburg \\
Germany\\
FAX: ++49/761/203-5967 \\
e-mail: jeti @ fdm.uni-freiburg.de
\clearpage

\begin {abstract}

A statistical test is presented to decide whether data are adequately
described by probabilistic functions of finite state Markov 
chains (''hidden Markov models'') as applied in the
analysis of ion channel data.
%
Particularly, the test can be used to decide whether a system obeys the
Markov condition.
Simulation studies are performed in order to investigate the sensitivity
of the proposed test against violations of the model assumptions.
The test can be applied analogously to Markov models.

\end {abstract}

\vskip 0.5cm
\vskip 0.5cm

%
\clearpage
\section* {I. Introduction}
Ion channels are large proteins located in the membranes of cells.
They serve for signal transmission and regulate the concentration
of ions in the cell. The channels open and close in a stochastic
manner dependent on external conditions like trans-membrane voltage
difference, concentration of ligands or mechanical stress.
In general the channels have several states in which they are closed,
resp. open. They might even possess open states with different conductivity.
The noisy current in the range of $pA$ through single channels can be 
measured by the 
patch clamp technique \cite{neher}.

Analyzing 
data from ion channels generally
relies on the assumption of a Markovian dynamics. This holds
for infering the number of channel states and mean dwell times 
 by fitting exponentials to dwell time histograms
\cite{kijima,colquhoun},
 for explicit modeling of low-pass filtered 
records by Markov models \cite{horn83,horn84}
and also for analyzing unfiltered records by hidden
Markov models \cite{chung90,chung91,fredkin92a,fredkin92b,becker,albertsen}.


In many cases, however, it is not evident from empirical data whether the
system actually obeys the Markov condition. 
For two reasons, this assumption has given rise to a lively discussion
\cite{korn,mcmanus,liebovitch89}.
On the one hand, the information about the validity of this
condition can provide valuable insight into the system under
investigation \cite{liebovitch87,sansom}.
On the other hand, conclusions drawn from a 
model which does not fit to the process that has
produced the data are very likely to lead to erroneous results.
Thus, it is desirable to test whether
the process is adequately described by the selected (hidden) Markov model.


We propose a test to perform this task.
It is based on the asymptotic distribution of the log-likelihood 
that holds if the model is valid. 
A deviation from the expected distribution provides a test
for the model. In order to evaluate the sensitivity of the proposed
test against 
a violation of the null hypothesis four simulation studies are 
performed where the assumption of an underlying hidden
Markov model is violated in various manners. A fifth 
simulation study shows that the test is also useful to estimate
the minimum number of states in a Markov model needed to be 
 compatible with the data.

The paper is organized as follows: In the next section, we briefly review 
the hidden Markov model. In Section III the
 test statistic is introduced. 
The power of test is evaluated by simulation studies in the
Section IV. As presented, the test applies to hidden Markov models, however, 
it can be applied analogously to Markov models.

\section* {II. Hidden Markov Models}

Hidden Markov models (HMM), introduced by \cite{baum66} and used in
diverse fields like speech recognition  \cite{rabiner} and ion
channel analysis, are generalizations of Markov models that allow to include 
observational noise. HMMs can be formulated in continuous-time and
discrete-time versions. Following \cite{chung91} we chose the
latter. The results also hold for  continuous-time models.


A stationary hidden Markov model is given by an unobservable  process
$ {\bf {X}_t} $ which can take one of $ s $ states for every point $t$
in time.
The probabilities for a change from a state $i$ to a state $j$ are described
by a time independent transition matrix $ (a_{ij})$ $ (i,j=1,\ \dots, \ s)$.
 Since each row
of the matrix is normalized to unity, the $ s \times s $ matrix $(a_{ij})$
has $ s (s-1) $ free parameters. The observations
 $ Y_t $  are determined by  the output probability densities 
 of each of the $ s $ states. These densities are described
 by parameter vectors $ {\bphi}_i $ $ (i=1,\ \dots, \ s)$. For example,
the density functions $ f (y, {\bphi}_i) $ can be given by Gaussian
distributions with different means and variances.

For ease of notation, the parameters of the hidden Markov model are
arranged in a single parameter vector  $ \btheta $.
Its dimension is denoted by $ r $. For example,
in the case of $ s $ states with Gaussian output probabilities the model has
 $ r = s (s-1) + 2s = s^2 + s $ parameters.

Given an observed time series $ {\bf{Y}}_{1 \ldots N}  = Y_1,\ \ldots, \ Y_N $
 of length N, the parameter vector $ \btheta $ can be estimated by a maximum
likelihood procedure \cite{baum70,levinson,qian}.
 For the calculation of the log-likelihood function
\beq
 L_N(\bY_{1 \ldots N},\btheta) = \log \ P[ \bY_{1 \ldots N }|\btheta ]
\eeq
 the so called forward
probabilities to find the system in state $i$ at time $t$ given the
data up to time $t$ are defined by :
\beq
 \alpha_i (t,\bY_{1 \ldots t},\btheta) = P[ X_t=i, \bY_{1 \ldots
             t}|\btheta ] \quad .
\eeq
They can be calculated using the recursion:
\beq
   \alpha_i (t,\bY_{1 \ldots t},\btheta) = \sum_{j=1}^s \, \alpha_j
(t-1,\bY_{1 \ldots t-1},\btheta) \, a_{ij} \, (\btheta) f(Y_t,\bphi_j)
\eeq
and lead to the log-likelihood function by:
\beq
    L_N(\bY_{1 \ldots N }|\btheta ) = \log \sum_{i=1}^s \, \alpha_i (N,\bY_{1
\ldots N},\btheta) \quad .
\eeq
%

An estimate $ {\hat {\btheta}_N} $ can be obtained by maximizing $
L_N(\bY_{1 \ldots N }|\btheta ) $ with respect to $ {\btheta} $
either by nonlinear optimization or by the Expectation -- Maximization
algorithm, i.e.  the Baum - Welsh reestimation formulae 
\cite{baum70,dempster}.
Here, all numerical calculations have been performed
by the latter method as described in \cite{levinson} since it 
 behaves numerically more stable than nonlinear optimization.
 For ease of notation we suppress
 the dependence of $ L_N(\bY_{1 \ldots N }|\btheta ) $ on
$\bY_{1 \ldots N}$ in the following.

\section* {III. The Test Statistic}

In this section 
we introduce the  statistic to test the adequacy of a given
hidden Markov model to describe an observed time series.

Under mild regularity conditions, the difference between the
maximum likelihood estimators 
$\hat{\btheta}_N $ and the true parameters $\btheta_0 $ are generally
believed due to central limit theorems to converge to a normal distribution
\beq \label{asy_normal}
  \sqrt{N}\, (\btheta_0 - \hat{\btheta}_N) \sim {\cal N} (0, \Sigma)
\eeq
with asymptotically :
\beq \label{asy_info}
  \frac{\partial^2}{\partial \theta_i \partial \theta_j} L_N(\hat{\btheta}_N)
	\rightarrow   - \frac{1}{N} \Sigma_{ij}^{-1}  \quad .
\eeq
%
%

This has been proven for independent random variables (see e.g.~\cite{cox}),
Markov models
\cite{billingsley} and hidden Markov models with discrete output 
probabilities \cite{baum66}. For hidden Markov models with continuous
output probabilities, up to now, the consistency of the
maximum likelihood estimators \cite{leroux}, the local asymptotic 
normality in the sense of Le Cam \cite{lecam,bickel} and the
asymptotic normality of maximum split data likelihood estimators 
has been shown \cite{ryden1}. The proof of asymptotic normality 
of the maximum likelihood estimators in hidden Markov models is
announced \cite{ryden2}.

Given the asymptotic normality of the estimators of Eqs.~(\ref{asy_normal}), 
the distribution of the maximum log-likelihood $L_N (\hat{\btheta}_N)$
that is itself a random variable can be derived by a Taylor expansion 
(see \cite{cox} for a detailed discussion) : 
\begin{eqnarray}
L_N(\btheta_0) & = & L_N(\hat{\btheta}_N) + 
   \frac{\partial}{\partial\, \theta_i} L_N(\hat{\btheta}_N)
			(\btheta_0 -\hat{\btheta}_N) + \nonumber \\
 &&  \frac{1}{2} (\btheta_0 - \hat{\btheta}_N) \frac{\partial^2}
   {\partial \theta_i \partial \theta_j} L_N(\hat{\btheta}_N)
	(\btheta_0 - \hat{\btheta}_N)  + 
		O(|\btheta_0 - \hat{\btheta}_N|^3) \quad.
\end{eqnarray}
%
%
The second term on the right hand side vanishes due to the estimation 
procedure. Neglecting higher order terms, 
solving for $2 (L(\hat{\btheta}_N) - L(\btheta_0) )$ and using
Eqs.~(\ref{asy_normal}) and (\ref{asy_info}) yields :
\beq \label {asymptotic}
   2 \, [L_N(\hat{\btheta}_N) - L_N(\btheta_0)] \sim \chi^2_r \quad .
\eeq 
%
%
%
%
%
%
%
%
%
%
This relation holds asymptotically if the model is specified correctly.
The number $N$ of data needed to reach the asymptotic regime depends
on the process. Simulation studies not presented here show that 
Eq.~(\ref{asymptotic}) holds if each transition between the states
has occurred at least $10$ times.

For the test 
we estimate $ \btheta_0 $ based on the whole
time series of length $N$ and denote this estimate by $ \hat{\btheta}_N $.
Then, the time series is divided in $K$ parts
of length $M=N/K$. For each the these parts we estimate the parameters
$\hat{\btheta}_M$ and evaluate the log-likelihoods
$L_M(\hat{\btheta}_M)$ and $L_M(\hat{\btheta}_N)$. Asymptotically, 
i.e.~for $N \rightarrow \infty$,
 $M\rightarrow \infty$, but $M/N\rightarrow 0$, the distribution 
of $2 \, [L_M(\hat{\btheta}_M) - L_M(\hat{\btheta}_N)] $ is given by:
\beq \label {asymptotic2}
   2 \, [L_M(\hat{\btheta}_M) - L_M(\hat{\btheta}_N)] \sim \chi^2_r \quad .
\eeq
By the proposed procedure we obtain $K$ samples of the $ \chi^2_r $
distribution if the model is valid.
In order to judge whether Eq.~(\ref{asymptotic2}) holds, we apply
the Kolmogorov -- Smirnov -- test for the consistency of an
empirical distribution with a proposed theoretical distribution
\cite{sachs}. The  Kolmogorov -- Smirnov -- test statistic is denoted by 
${\cal {Z}}$ in the following.

\section* {IV. Evaluation of the power of the test} \label{power}
In this section, we evaluate the power of the above proposed test,
i.e. we investigate the sensitivity of the test against a violation
of the null hypothesis that the data were produced by a hidden
Markov model.
Of course, it is not possible to consider
all imaginable alternative hypotheses. One has to restrict oneself to a
reasonable class of alternative hypotheses.
We choose four alternative hypotheses that violate the model assumptions:

\begin {itemize}
  \item Nonstationary transition probabilities
  \item Dwell time dependent transition probabilities
  \item A fractal model
  \item Refractory time
\end {itemize}
Finally, we show that the proposed test enables to estimate
the smallest number of states of the Markov process compatible with the data.
\vskip 0.4cm

\subsection* {A. Model definition}
%

%
In order to evaluate the power of the test numerically
we chose a hidden Markov model with three
states and Gaussian output probability functions representing e.g. one ion 
channel with two different conductance levels.
The transition matrix $A$ is given by:
\beq \label{model}
A = \left(
\begin{array}{ccc}
   0.90  &  0.05  &  0.05    \\
   0.06   & 0.92  &  0.02     \\
   0.03   & 0.02   & 0.95
\end{array}
\right)
\eeq
The means and the variances of the gaussian output probability functions
were chosen to be:
\beq \label{model1}
\begin{array}{ccc}
   \mu_1 = 0.0 & \quad & \sigma_1^2 = 0.1 \\
   \mu_2 = 1.0 & \quad & \sigma_2^2 = 0.1 \\
   \mu_3 = 2.0 & \quad & \sigma_3^2 = 0.1
\end{array}
\eeq
The dimension $r$ of the parameter vector $\btheta$ is 12.
We simulated time series of length $N=150.000 $ and divided it into $K=150$
time series of length $M=1000$ to perform the test. 
%
%
To apply the test the resulting time series must be long enough
for the asymptotic results to be valid. 
%
%
If the off - diagonal elements of the 
transition matrix are of similar magnitude, as a rule of thumb,  
this condition is met, if the time series have a length of at least : 
\beq
  M= 10 \, s \, \tau_{max}  
\eeq
with $s$ the number of states and $ \tau_{max} $ the largest dwell time.
For the chosen model, the dwell times are  $ 10,\, 12.5,$ resp. $20$ units of
time.
\marginpar{please locate Fig. \ref{C} around here}

Fig. \ref{C} shows the expected cumulative $\chi^2_{12}$ distribution 
according to Eq.~(\ref{asymptotic2}) and the empirical cumulative
distribution for the chosen process. 
It indicates a good qualitative agreement of the two distributions.
In order to quantify this, we counted for 200 realizations of the
process the number of cases where the hypothesis of consistency of the
two distributions were rejected by the Kolmogorov - Smirnov - test
at a significance level of 5 \%. 
This results in actual rejection rate of 4.5 \%,
indicating that the asymptotic regime is reached for the chosen situation.

\subsection* {B. Power of the test}

To investigate the power of a test, usually, for different degrees of
violation of the null hypothesis in the order of thousand
 time series are realized,
the test is performed and the fraction of rejected null hypothesis
given a certain significance level $\alpha$ is calculated  in dependence
 of the degree of violation.
%
%
However, this procedure to evaluate the power requires an enormous
computational effort to obtain a good approximation of the underlying smooth
behavior since 
for the chosen model and number of data the maximization of the
log-likelihood for a single time series requires ca.~45 min.~on an 
IBM 6000 RISC workstation.
Therefore, we choose another way to display the 
power of the test. Instead of counting the simulation runs with
rejected null hypothesis we
average the test statistic of $10$ realizations for each degree of
violation of the null hypothesis to approximate the smooth curve.
This procedure estimates the mean of the distribution of the
test statistic for the alternative hypotheses. 
Simulation studies show that these distributions of the test statistic
are symmetric and that their variance is rather constant. Therefore, the
mean calculated here corresponds to the median of the distributions 
and is related monotonically to the fraction calculated usually.
Thus, 
%
this
procedure yields essentially the same information as the canonical method
but requires only one per cent of
computational effort.

\vskip 0.5cm
We now discuss the different simulation studies to evaluate the power of the
proposed test.


\begin {itemize}
  \item  Nonstationary transition probabilities              \\
       In order to investigate the sensitivity against violations of the
       stationarity assumption, 
       nonstationarity of the transition probability of the first
       state is introduced by :
       \begin{eqnarray}
         \tilde{a}_{11}(t) & = & a_{11} - \frac{(s-1)\, \nu\, t}{N} \\
         \tilde{a}_{1j}(t) & = & a_{1j} + \frac{\nu\, t}{N} ,\quad (j= 2,
               \ldots , s)    \qquad ,
       \end{eqnarray}

       where $s$ again denotes the number of states. This
       time dependency of the transition probabilities causes a decreasing
       dwell time of the first state.
       The drift rate $\nu$ serves as the parameter for the null
       hypothesis  violation.
\marginpar{please locate Fig. \ref{nonstat} around here}
       As outlined above, we judge the performance of the test by
       averaging the test statistic of ten simulations for every degree of
       the null hypothesis violation.  Fig. \ref{nonstat} shows the
       averaged test statistic ${\cal{Z}}$ of the Kolmogorov - Smirnov -
       test with increasing violation
       of the null  hypothesis and the $1 \%$, resp. $0.1 \%$ levels of
       significance. A change of $ 10\% $ over the whole observation time 
       in the dwell probability
       of one of three states is detectable by the proposed test.
   \item Dwell time dependent transition probabilities         \\
        The Markov condition, stating that the transition probabilities
        between the states do not depend on the time already spent
        in the states 
        is violated  by increasing
        the probability to leave any state proportional to the time
        $ t_{in} $ already spent in the state. The proportionality
        constant $ \gamma $ parameterizes the violation of the null
        hypothesis.
      \begin{eqnarray}
        \tilde{a}_{ii}(t_{in}) & = & a_{ii} - (s-1) \, \gamma t_{in} \\
	\tilde{a}_{ij}(t_{in}) & = & a_{ij} + \gamma \,t_{in},
		\quad (i \ne j)
      \end{eqnarray}
\marginpar{please locate Fig. \ref{nonmark} around here}
       Fig. \ref{nonmark} shows the result of the simulation. A change of
       more than one per cent per time step for the dwell
       probabilities leads to the rejection of the  hypothesis that the time
       series was generated by a Markov process. During the simulation, 
       it was controlled that
       the condition $ 0 < \tilde{a}_{ij}(t_{in}) < 1 $ was not violated.

   \item A fractal model\\
   Another possibility to  violate the Markov condition is given by the
   fractal models \cite{liebovitch87}. For these models,
  the dwell probability increases with the time $t_{in} $ already spent 
  in the state.
   The transition probabilities of a fractal model are given by:
      \begin{eqnarray}
       \tilde{a}_{ii}(t_{in}) & = & 1 - (1 - a_{ii})\, t_{in}^{1-D}  \\
       \tilde{a}_{ij}(t_{in}) & = & a_{ij} \, t_{in}^{1-D}   
\qquad i \ne j \quad ,
      \end{eqnarray}
\marginpar{please locate Fig. \ref{frac} around here}
	where  $D$ is the fractal dimension which parameterizes the
  violation of the Markov condition. For $D=1$ the Markov model results.
    The result of the simulation in  Fig. \ref{frac} reveals that for the 
   given model a fractal dimension of e.g.~$1.1$ will lead to a 
   rejection of a Markovian process. On the other hand, a 
   dimension larger than $1.1$ can be excluded if the test does not
  rejected the model. 
  \item Refractory time\\
       Finally, the Markov condition is violated by introducing
       a refractory time, i.e.~a minimal time that the process has to spend
       in a state. To simulate such processes we used the  model
       according
       to Eq.~(\ref{model}) but forced the state to stay for the time
       $ \tau_{\mbox{ref }} $ before the dynamics were applied.
\marginpar{please locate Fig. \ref{refrac} around here}
       Fig. \ref{refrac} displays the results. Note that the dwell times of
       the chosen model were $10, \, 12.5 $, resp. $20$ units of time, so
       that the considered type of violation
       is only detectable if it amounts to $ 50 \%$ of the shortest dwell time.

\end {itemize}
In summary, the test enables a detection of different types of
violations of the Markov condition.

\subsection* {C. Estimating the minimum number of states }
%
So far, the number of states $s$ of the Markov process was assumed
to be known. Since the number of assumed states $\hat{s}$ determines
the degrees of freedom $r$ of the model, the proposed test can
be applied to infer the number of states of the process under investigation.
This is done by comparing the left hand side of Eq.~(\ref{asymptotic2})
with the $\chi^2_{\hat{r}}$ distribution with degrees of freedom $\hat{r}$
corresponding to the assumed model, e.g.~in the case of a Gaussian model:
 $\hat{r}=\hat{s}^2 + \hat{s}$.
%
\marginpar{please locate Fig. \ref{states} around here}
Fig. \ref{states} displays the results.
Hidden Markov models with increasing number of
states are fitted to data from the model Eq.~(\ref{model}) with three
states. The test enables a determination of the (correct) smallest
number of states that can describe the process. Note that models
with more than three states are also detected as being consistent
with the data. 

\section* {V. Discussion}

Markov and hidden Markov models are increasingly used in the analysis
 of patch clamp
ion channel data. In many cases their adequacy for a given system has
been assumed, but not tested using empirical data.
If a record is a realization of a hidden Markov process, the
asymptotic distribution of the log-likelihood is a $\chi^2_r$ distribution,
its number of degrees of freedom $r$ being given by the number of model
parameters. Thus, a test
for the consistency of the empirical distribution of a fitted model with the
theoretical distribution provides a test whether the time series may be
considered as a realization of a hidden Markov process. 

Based on the asymptotic distribution of the log-likelihood, we have
 introduced 
 such a test. The test is analogously applicable to test
Markov models. In order to investigate
how sensitive the test is to detect a violation of the assumed model, we
performed four simulation studies where we modified
a hidden Markov process continuously in different ways to become
nonmarkovian.
The sensitivity of the proposed test depends on how
the model assumption of a stationary Markov process is violated:
The test has shown to be very sensitive if the violation results from
drifting transition probabilities, from dwell time dependent transition
probabilities or a fractal model. 
For example, a fractal dimension of $1.7$ as reported in 
\cite{liebovitch87} would lead to a highly significant rejection of the
Markov model used in the simulation study. 
The test is less sensitive to detect
refractory times in the system that retard the beginning of the
Markovian dynamics.

Furthermore, the proposed test can be used to estimate the minimum number of
states in the Markov process necessary to describe the data.

In applications, performing simulation studies as presented
will reveal which degree of violation of
the model assumptions is consistent with the fitted model and which degrees
of violation can be excluded if the model can not be rejected.

The test is suited for analyzing data recorded
under steady state conditions like in the case of ligand dependent
ion channels. For voltage dependent channels where numerous
trials for a certain pulse protocol are recorded these single 
trials determine the length $ M $ in the proposed test. 
Further, it can help to decide whether observed changes 
in inactivation dynamics \cite{cannon} are consistent
with statistical fluctuations in a fitted model or have to be treated
explicitly as modal gating between two different dynamics.

\section* {Acknowledgements}
%
 We would like to thank A. Wilts and U.-P. Hansen
 for  valuable comments on an earlier version of this manuscript.

\clearpage

\section{Figure Captions}

\begin{description}

 \item [Fig. 1:] The empirical cumulative distribution
   of $ 2 \, [L_M(\hat{\btheta}_M) - L_M(\hat{\btheta}_N)] $ (solid line)
     and the expected cumulative $\chi^2_{12}$ distribution (dotted line) for 
     the process defined by Eqs.~(\ref {model}) and (\ref {model1}). 


  \item [Fig. 2:] The effect of drifting transition probabilities.
         Shown is the averaged test statistic ${\cal {Z}}$ 
         of the Kolmogorov - Smirnov test for
         increasing  drift rates $\nu$.
	The $ 1\% $ and the $0.1 \% $ significance levels are
         marked.

  \item [Fig. 3:] The effect of dwell time dependent transition
         probabilities. Shown is the averaged test statistic ${\cal{Z}}$ 
         for increasing  degrees $\gamma$ of the null
         hypothesis  violation.
        The $ 1\% $ and the $0.1 \% $ significance levels are marked.

  \item [Fig. 4:] Violation of the Markov condition by a fractal model.
         Shown is the averaged test statistic ${\cal{Z}}$ for
         increasing fractal dimension $D$.

  \item [Fig. 5:] The effect of refractory time. Shown is the averaged
        test statistic ${\cal{Z}}$ for increasing refractory times
        $\tau_{\mbox{ref }}$.

  \item [Fig. 6:] Determining the number of states. Shown is
        averaged test statistic ${\cal{Z}}$ for hidden
        Markov models with different number of states $\hat{s}$ applied to 
        time series that were generated by a Hidden Markov Model with three
        states.
The $ 1\% $ and the $0.1 \% $ significance levels are
         marked.

\end{description}

\clearpage

\begin{figure} [t]
    \input{C.tex  }
    \caption{\label{C}    }
\end {figure}
\clearpage


\clearpage
\begin{figure} [t]
\setlength{\unitlength}{0.240900pt}
\ifx\plotpoint\undefined\newsavebox{\plotpoint}\fi
\begin{picture}(1500,900)(0,0)
\font\gnuplot=cmr10 at 10pt
\gnuplot
\sbox{\plotpoint}{\rule[-0.200pt]{0.400pt}{0.400pt}}%
\put(220.0,113.0){\rule[-0.200pt]{292.934pt}{0.400pt}}
\put(220.0,113.0){\rule[-0.200pt]{0.400pt}{184.048pt}}
\put(220.0,113.0){\rule[-0.200pt]{4.818pt}{0.400pt}}
\put(198,113){\makebox(0,0)[r]{0}}
\put(1416.0,113.0){\rule[-0.200pt]{4.818pt}{0.400pt}}
\put(220.0,240.0){\rule[-0.200pt]{4.818pt}{0.400pt}}
\put(198,240){\makebox(0,0)[r]{0.05}}
\put(1416.0,240.0){\rule[-0.200pt]{4.818pt}{0.400pt}}
\put(220.0,368.0){\rule[-0.200pt]{4.818pt}{0.400pt}}
\put(198,368){\makebox(0,0)[r]{0.1}}
\put(1416.0,368.0){\rule[-0.200pt]{4.818pt}{0.400pt}}
\put(220.0,495.0){\rule[-0.200pt]{4.818pt}{0.400pt}}
\put(198,495){\makebox(0,0)[r]{0.15}}
\put(1416.0,495.0){\rule[-0.200pt]{4.818pt}{0.400pt}}
\put(220.0,622.0){\rule[-0.200pt]{4.818pt}{0.400pt}}
\put(198,622){\makebox(0,0)[r]{0.2}}
\put(1416.0,622.0){\rule[-0.200pt]{4.818pt}{0.400pt}}
\put(220.0,750.0){\rule[-0.200pt]{4.818pt}{0.400pt}}
\put(198,750){\makebox(0,0)[r]{0.25}}
\put(1416.0,750.0){\rule[-0.200pt]{4.818pt}{0.400pt}}
\put(220.0,877.0){\rule[-0.200pt]{4.818pt}{0.400pt}}
\put(198,877){\makebox(0,0)[r]{0.3}}
\put(1416.0,877.0){\rule[-0.200pt]{4.818pt}{0.400pt}}
\put(220.0,113.0){\rule[-0.200pt]{0.400pt}{4.818pt}}
\put(220,68){\makebox(0,0){0}}
\put(220.0,857.0){\rule[-0.200pt]{0.400pt}{4.818pt}}
\put(372.0,113.0){\rule[-0.200pt]{0.400pt}{4.818pt}}
\put(372,68){\makebox(0,0){0.01}}
\put(372.0,857.0){\rule[-0.200pt]{0.400pt}{4.818pt}}
\put(524.0,113.0){\rule[-0.200pt]{0.400pt}{4.818pt}}
\put(524,68){\makebox(0,0){0.02}}
\put(524.0,857.0){\rule[-0.200pt]{0.400pt}{4.818pt}}
\put(676.0,113.0){\rule[-0.200pt]{0.400pt}{4.818pt}}
\put(676,68){\makebox(0,0){0.03}}
\put(676.0,857.0){\rule[-0.200pt]{0.400pt}{4.818pt}}
\put(828.0,113.0){\rule[-0.200pt]{0.400pt}{4.818pt}}
\put(828,68){\makebox(0,0){0.04}}
\put(828.0,857.0){\rule[-0.200pt]{0.400pt}{4.818pt}}
\put(980.0,113.0){\rule[-0.200pt]{0.400pt}{4.818pt}}
\put(980,68){\makebox(0,0){0.05}}
\put(980.0,857.0){\rule[-0.200pt]{0.400pt}{4.818pt}}
\put(1132.0,113.0){\rule[-0.200pt]{0.400pt}{4.818pt}}
\put(1132,68){\makebox(0,0){0.06}}
\put(1132.0,857.0){\rule[-0.200pt]{0.400pt}{4.818pt}}
\put(1284.0,113.0){\rule[-0.200pt]{0.400pt}{4.818pt}}
\put(1284,68){\makebox(0,0){0.07}}
\put(1284.0,857.0){\rule[-0.200pt]{0.400pt}{4.818pt}}
\put(1436.0,113.0){\rule[-0.200pt]{0.400pt}{4.818pt}}
\put(1436,68){\makebox(0,0){0.08}}
\put(1436.0,857.0){\rule[-0.200pt]{0.400pt}{4.818pt}}
\put(220.0,113.0){\rule[-0.200pt]{292.934pt}{0.400pt}}
\put(1436.0,113.0){\rule[-0.200pt]{0.400pt}{184.048pt}}
\put(220.0,877.0){\rule[-0.200pt]{292.934pt}{0.400pt}}
\put(45,495){\makebox(0,0){${\cal Z}$ }}
\put(828,23){\makebox(0,0){$\nu$}}
\put(1284,472){\makebox(0,0)[l]{1 \%}}
\put(1284,548){\makebox(0,0)[l]{0.1 \%}}
\put(220.0,113.0){\rule[-0.200pt]{0.400pt}{184.048pt}}
\sbox{\plotpoint}{\rule[-0.400pt]{0.800pt}{0.800pt}}%
\put(220,317){\usebox{\plotpoint}}
\multiput(220.00,315.09)(0.747,-0.502){95}{\rule{1.392pt}{0.121pt}}
\multiput(220.00,315.34)(73.111,-51.000){2}{\rule{0.696pt}{0.800pt}}
\multiput(296.00,267.41)(3.098,0.509){19}{\rule{4.877pt}{0.123pt}}
\multiput(296.00,264.34)(65.878,13.000){2}{\rule{2.438pt}{0.800pt}}
\multiput(372.00,280.41)(3.384,0.511){17}{\rule{5.267pt}{0.123pt}}
\multiput(372.00,277.34)(65.069,12.000){2}{\rule{2.633pt}{0.800pt}}
\multiput(448.00,289.08)(3.384,-0.511){17}{\rule{5.267pt}{0.123pt}}
\multiput(448.00,289.34)(65.069,-12.000){2}{\rule{2.633pt}{0.800pt}}
\multiput(524.00,280.41)(1.549,0.504){43}{\rule{2.632pt}{0.121pt}}
\multiput(524.00,277.34)(70.537,25.000){2}{\rule{1.316pt}{0.800pt}}
\multiput(600.00,302.08)(3.098,-0.509){19}{\rule{4.877pt}{0.123pt}}
\multiput(600.00,302.34)(65.878,-13.000){2}{\rule{2.438pt}{0.800pt}}
\multiput(677.41,291.00)(0.501,0.672){145}{\rule{0.121pt}{1.274pt}}
\multiput(674.34,291.00)(76.000,99.356){2}{\rule{0.800pt}{0.637pt}}
\multiput(752.00,394.41)(1.488,0.504){45}{\rule{2.538pt}{0.121pt}}
\multiput(752.00,391.34)(70.731,26.000){2}{\rule{1.269pt}{0.800pt}}
\multiput(828.00,420.41)(0.747,0.502){95}{\rule{1.392pt}{0.121pt}}
\multiput(828.00,417.34)(73.111,51.000){2}{\rule{0.696pt}{0.800pt}}
\multiput(904.00,468.08)(3.098,-0.509){19}{\rule{4.877pt}{0.123pt}}
\multiput(904.00,468.34)(65.878,-13.000){2}{\rule{2.438pt}{0.800pt}}
\multiput(980.00,458.41)(0.603,0.502){119}{\rule{1.165pt}{0.121pt}}
\multiput(980.00,455.34)(73.582,63.000){2}{\rule{0.583pt}{0.800pt}}
\multiput(1056.00,521.41)(1.488,0.504){45}{\rule{2.538pt}{0.121pt}}
\multiput(1056.00,518.34)(70.731,26.000){2}{\rule{1.269pt}{0.800pt}}
\multiput(1132.00,547.41)(0.747,0.502){95}{\rule{1.392pt}{0.121pt}}
\multiput(1132.00,544.34)(73.111,51.000){2}{\rule{0.696pt}{0.800pt}}
\multiput(1208.00,595.09)(1.488,-0.504){45}{\rule{2.538pt}{0.121pt}}
\multiput(1208.00,595.34)(70.731,-26.000){2}{\rule{1.269pt}{0.800pt}}
\multiput(1285.41,571.00)(0.501,0.506){145}{\rule{0.121pt}{1.011pt}}
\multiput(1282.34,571.00)(76.000,74.903){2}{\rule{0.800pt}{0.505pt}}
\multiput(1361.41,648.00)(0.501,0.924){145}{\rule{0.121pt}{1.674pt}}
\multiput(1358.34,648.00)(76.000,136.526){2}{\rule{0.800pt}{0.837pt}}
\sbox{\plotpoint}{\rule[-0.200pt]{0.400pt}{0.400pt}}%
\put(220,444){\usebox{\plotpoint}}
\multiput(220,444)(20.756,0.000){59}{\usebox{\plotpoint}}
\put(1436,444){\usebox{\plotpoint}}
\put(220,520){\usebox{\plotpoint}}
\multiput(220,520)(20.756,0.000){59}{\usebox{\plotpoint}}
\put(1436,520){\usebox{\plotpoint}}
\end{picture}
    \caption{\label{nonstat}    }
\end {figure}

\clearpage
\begin{figure} [t]
\setlength{\unitlength}{0.240900pt}
\ifx\plotpoint\undefined\newsavebox{\plotpoint}\fi
\begin{picture}(1500,900)(0,0)
\font\gnuplot=cmr10 at 10pt
\gnuplot
\sbox{\plotpoint}{\rule[-0.200pt]{0.400pt}{0.400pt}}%
\put(220.0,113.0){\rule[-0.200pt]{0.400pt}{184.048pt}}
\put(220.0,113.0){\rule[-0.200pt]{4.818pt}{0.400pt}}
\put(198,113){\makebox(0,0)[r]{0.05}}
\put(1416.0,113.0){\rule[-0.200pt]{4.818pt}{0.400pt}}
\put(220.0,304.0){\rule[-0.200pt]{4.818pt}{0.400pt}}
\put(198,304){\makebox(0,0)[r]{0.1}}
\put(1416.0,304.0){\rule[-0.200pt]{4.818pt}{0.400pt}}
\put(220.0,495.0){\rule[-0.200pt]{4.818pt}{0.400pt}}
\put(198,495){\makebox(0,0)[r]{0.15}}
\put(1416.0,495.0){\rule[-0.200pt]{4.818pt}{0.400pt}}
\put(220.0,686.0){\rule[-0.200pt]{4.818pt}{0.400pt}}
\put(198,686){\makebox(0,0)[r]{0.2}}
\put(1416.0,686.0){\rule[-0.200pt]{4.818pt}{0.400pt}}
\put(220.0,877.0){\rule[-0.200pt]{4.818pt}{0.400pt}}
\put(198,877){\makebox(0,0)[r]{0.25}}
\put(1416.0,877.0){\rule[-0.200pt]{4.818pt}{0.400pt}}
\put(220.0,113.0){\rule[-0.200pt]{0.400pt}{4.818pt}}
\put(220,68){\makebox(0,0){0}}
\put(220.0,857.0){\rule[-0.200pt]{0.400pt}{4.818pt}}
\put(372.0,113.0){\rule[-0.200pt]{0.400pt}{4.818pt}}
\put(372,68){\makebox(0,0){0.001}}
\put(372.0,857.0){\rule[-0.200pt]{0.400pt}{4.818pt}}
\put(524.0,113.0){\rule[-0.200pt]{0.400pt}{4.818pt}}
\put(524,68){\makebox(0,0){0.002}}
\put(524.0,857.0){\rule[-0.200pt]{0.400pt}{4.818pt}}
\put(676.0,113.0){\rule[-0.200pt]{0.400pt}{4.818pt}}
\put(676,68){\makebox(0,0){0.003}}
\put(676.0,857.0){\rule[-0.200pt]{0.400pt}{4.818pt}}
\put(828.0,113.0){\rule[-0.200pt]{0.400pt}{4.818pt}}
\put(828,68){\makebox(0,0){0.004}}
\put(828.0,857.0){\rule[-0.200pt]{0.400pt}{4.818pt}}
\put(980.0,113.0){\rule[-0.200pt]{0.400pt}{4.818pt}}
\put(980,68){\makebox(0,0){0.005}}
\put(980.0,857.0){\rule[-0.200pt]{0.400pt}{4.818pt}}
\put(1132.0,113.0){\rule[-0.200pt]{0.400pt}{4.818pt}}
\put(1132,68){\makebox(0,0){0.006}}
\put(1132.0,857.0){\rule[-0.200pt]{0.400pt}{4.818pt}}
\put(1284.0,113.0){\rule[-0.200pt]{0.400pt}{4.818pt}}
\put(1284,68){\makebox(0,0){0.007}}
\put(1284.0,857.0){\rule[-0.200pt]{0.400pt}{4.818pt}}
\put(1436.0,113.0){\rule[-0.200pt]{0.400pt}{4.818pt}}
\put(1436,68){\makebox(0,0){0.008}}
\put(1436.0,857.0){\rule[-0.200pt]{0.400pt}{4.818pt}}
\put(220.0,113.0){\rule[-0.200pt]{292.934pt}{0.400pt}}
\put(1436.0,113.0){\rule[-0.200pt]{0.400pt}{184.048pt}}
\put(220.0,877.0){\rule[-0.200pt]{292.934pt}{0.400pt}}
\put(45,495){\makebox(0,0){${\cal Z}$ }}
\put(828,23){\makebox(0,0){$\gamma$}}
\put(1284,445){\makebox(0,0)[l]{1 \%}}
\put(1284,560){\makebox(0,0)[l]{0.1 \%}}
\put(220.0,113.0){\rule[-0.200pt]{0.400pt}{184.048pt}}
\sbox{\plotpoint}{\rule[-0.400pt]{0.800pt}{0.800pt}}%
\put(220,208){\usebox{\plotpoint}}
\multiput(220.00,209.41)(1.955,0.505){33}{\rule{3.240pt}{0.122pt}}
\multiput(220.00,206.34)(69.275,20.000){2}{\rule{1.620pt}{0.800pt}}
\multiput(296.00,229.41)(0.499,0.501){145}{\rule{1.000pt}{0.121pt}}
\multiput(296.00,226.34)(73.924,76.000){2}{\rule{0.500pt}{0.800pt}}
\multiput(373.41,304.00)(0.501,0.758){145}{\rule{0.121pt}{1.411pt}}
\multiput(370.34,304.00)(76.000,112.072){2}{\rule{0.800pt}{0.705pt}}
\multiput(448.00,417.09)(2.063,-0.506){31}{\rule{3.400pt}{0.122pt}}
\multiput(448.00,417.34)(68.943,-19.000){2}{\rule{1.700pt}{0.800pt}}
\multiput(524.00,401.41)(1.008,0.503){69}{\rule{1.800pt}{0.121pt}}
\multiput(524.00,398.34)(72.264,38.000){2}{\rule{0.900pt}{0.800pt}}
\multiput(600.00,439.41)(0.778,0.502){91}{\rule{1.441pt}{0.121pt}}
\multiput(600.00,436.34)(73.010,49.000){2}{\rule{0.720pt}{0.800pt}}
\multiput(676.00,485.09)(1.284,-0.503){53}{\rule{2.227pt}{0.121pt}}
\multiput(676.00,485.34)(71.378,-30.000){2}{\rule{1.113pt}{0.800pt}}
\multiput(752.00,458.41)(1.008,0.503){69}{\rule{1.800pt}{0.121pt}}
\multiput(752.00,455.34)(72.264,38.000){2}{\rule{0.900pt}{0.800pt}}
\multiput(828.00,496.40)(5.453,0.520){9}{\rule{7.800pt}{0.125pt}}
\multiput(828.00,493.34)(59.811,8.000){2}{\rule{3.900pt}{0.800pt}}
\multiput(904.00,504.41)(1.689,0.505){39}{\rule{2.843pt}{0.122pt}}
\multiput(904.00,501.34)(70.098,23.000){2}{\rule{1.422pt}{0.800pt}}
\multiput(980.00,524.08)(5.453,-0.520){9}{\rule{7.800pt}{0.125pt}}
\multiput(980.00,524.34)(59.811,-8.000){2}{\rule{3.900pt}{0.800pt}}
\multiput(1056.00,519.41)(0.499,0.501){145}{\rule{1.000pt}{0.121pt}}
\multiput(1056.00,516.34)(73.924,76.000){2}{\rule{0.500pt}{0.800pt}}
\multiput(1132.00,595.41)(2.475,0.507){25}{\rule{4.000pt}{0.122pt}}
\multiput(1132.00,592.34)(67.698,16.000){2}{\rule{2.000pt}{0.800pt}}
\multiput(1208.00,608.09)(2.063,-0.506){31}{\rule{3.400pt}{0.122pt}}
\multiput(1208.00,608.34)(68.943,-19.000){2}{\rule{1.700pt}{0.800pt}}
\multiput(1285.41,591.00)(0.501,0.599){145}{\rule{0.121pt}{1.158pt}}
\multiput(1282.34,591.00)(76.000,88.597){2}{\rule{0.800pt}{0.579pt}}
\multiput(1360.00,683.41)(1.689,0.505){39}{\rule{2.843pt}{0.122pt}}
\multiput(1360.00,680.34)(70.098,23.000){2}{\rule{1.422pt}{0.800pt}}
\sbox{\plotpoint}{\rule[-0.200pt]{0.400pt}{0.400pt}}%
\put(220,419){\usebox{\plotpoint}}
\multiput(220,419)(20.756,0.000){59}{\usebox{\plotpoint}}
\put(1436,419){\usebox{\plotpoint}}
\put(220,533){\usebox{\plotpoint}}
\multiput(220,533)(20.756,0.000){59}{\usebox{\plotpoint}}
\put(1436,533){\usebox{\plotpoint}}
\end{picture}
   \caption{\label{nonmark}   }
\end {figure}

\clearpage
\begin{figure} [t]
\setlength{\unitlength}{0.240900pt}
\ifx\plotpoint\undefined\newsavebox{\plotpoint}\fi
\begin{picture}(1500,900)(0,0)
\font\gnuplot=cmr10 at 10pt
\gnuplot
\sbox{\plotpoint}{\rule[-0.200pt]{0.400pt}{0.400pt}}%
\put(220.0,164.0){\rule[-0.200pt]{4.818pt}{0.400pt}}
\put(198,164){\makebox(0,0)[r]{0.06}}
\put(1416.0,164.0){\rule[-0.200pt]{4.818pt}{0.400pt}}
\put(220.0,266.0){\rule[-0.200pt]{4.818pt}{0.400pt}}
\put(198,266){\makebox(0,0)[r]{0.08}}
\put(1416.0,266.0){\rule[-0.200pt]{4.818pt}{0.400pt}}
\put(220.0,368.0){\rule[-0.200pt]{4.818pt}{0.400pt}}
\put(198,368){\makebox(0,0)[r]{0.1}}
\put(1416.0,368.0){\rule[-0.200pt]{4.818pt}{0.400pt}}
\put(220.0,470.0){\rule[-0.200pt]{4.818pt}{0.400pt}}
\put(198,470){\makebox(0,0)[r]{0.12}}
\put(1416.0,470.0){\rule[-0.200pt]{4.818pt}{0.400pt}}
\put(220.0,571.0){\rule[-0.200pt]{4.818pt}{0.400pt}}
\put(198,571){\makebox(0,0)[r]{0.14}}
\put(1416.0,571.0){\rule[-0.200pt]{4.818pt}{0.400pt}}
\put(220.0,673.0){\rule[-0.200pt]{4.818pt}{0.400pt}}
\put(198,673){\makebox(0,0)[r]{0.16}}
\put(1416.0,673.0){\rule[-0.200pt]{4.818pt}{0.400pt}}
\put(220.0,775.0){\rule[-0.200pt]{4.818pt}{0.400pt}}
\put(198,775){\makebox(0,0)[r]{0.18}}
\put(1416.0,775.0){\rule[-0.200pt]{4.818pt}{0.400pt}}
\put(220.0,877.0){\rule[-0.200pt]{4.818pt}{0.400pt}}
\put(198,877){\makebox(0,0)[r]{0.2}}
\put(1416.0,877.0){\rule[-0.200pt]{4.818pt}{0.400pt}}
\put(220.0,113.0){\rule[-0.200pt]{0.400pt}{4.818pt}}
\put(220,68){\makebox(0,0){1}}
\put(220.0,857.0){\rule[-0.200pt]{0.400pt}{4.818pt}}
\put(463.0,113.0){\rule[-0.200pt]{0.400pt}{4.818pt}}
\put(463,68){\makebox(0,0){1.02}}
\put(463.0,857.0){\rule[-0.200pt]{0.400pt}{4.818pt}}
\put(706.0,113.0){\rule[-0.200pt]{0.400pt}{4.818pt}}
\put(706,68){\makebox(0,0){1.04}}
\put(706.0,857.0){\rule[-0.200pt]{0.400pt}{4.818pt}}
\put(950.0,113.0){\rule[-0.200pt]{0.400pt}{4.818pt}}
\put(950,68){\makebox(0,0){1.06}}
\put(950.0,857.0){\rule[-0.200pt]{0.400pt}{4.818pt}}
\put(1193.0,113.0){\rule[-0.200pt]{0.400pt}{4.818pt}}
\put(1193,68){\makebox(0,0){1.08}}
\put(1193.0,857.0){\rule[-0.200pt]{0.400pt}{4.818pt}}
\put(1436.0,113.0){\rule[-0.200pt]{0.400pt}{4.818pt}}
\put(1436,68){\makebox(0,0){1.1}}
\put(1436.0,857.0){\rule[-0.200pt]{0.400pt}{4.818pt}}
\put(220.0,113.0){\rule[-0.200pt]{292.934pt}{0.400pt}}
\put(1436.0,113.0){\rule[-0.200pt]{0.400pt}{184.048pt}}
\put(220.0,877.0){\rule[-0.200pt]{292.934pt}{0.400pt}}
\put(45,495){\makebox(0,0){${\cal Z}$ }}
\put(828,23){\makebox(0,0){$D$}}
\put(1314,546){\makebox(0,0)[l]{1 \%}}
\put(1314,699){\makebox(0,0)[l]{0.1 \%}}
\put(220.0,113.0){\rule[-0.200pt]{0.400pt}{184.048pt}}
\sbox{\plotpoint}{\rule[-0.400pt]{0.800pt}{0.800pt}}%
\put(220,164){\usebox{\plotpoint}}
\multiput(221.41,164.00)(0.501,0.627){237}{\rule{0.121pt}{1.203pt}}
\multiput(218.34,164.00)(122.000,150.503){2}{\rule{0.800pt}{0.602pt}}
\multiput(342.00,318.41)(1.195,0.502){95}{\rule{2.098pt}{0.121pt}}
\multiput(342.00,315.34)(116.645,51.000){2}{\rule{1.049pt}{0.800pt}}
\multiput(463.00,369.41)(1.005,0.502){115}{\rule{1.800pt}{0.121pt}}
\multiput(463.00,366.34)(118.264,61.000){2}{\rule{0.900pt}{0.800pt}}
\multiput(586.41,429.00)(0.501,0.632){235}{\rule{0.121pt}{1.212pt}}
\multiput(583.34,429.00)(121.000,150.485){2}{\rule{0.800pt}{0.606pt}}
\multiput(706.00,583.41)(4.280,0.508){23}{\rule{6.707pt}{0.122pt}}
\multiput(706.00,580.34)(108.080,15.000){2}{\rule{3.353pt}{0.800pt}}
\multiput(828.00,598.41)(2.499,0.504){43}{\rule{4.104pt}{0.121pt}}
\multiput(828.00,595.34)(113.482,25.000){2}{\rule{2.052pt}{0.800pt}}
\multiput(950.00,623.41)(2.972,0.505){35}{\rule{4.810pt}{0.122pt}}
\multiput(950.00,620.34)(111.018,21.000){2}{\rule{2.405pt}{0.800pt}}
\multiput(1071.00,644.41)(0.754,0.501){155}{\rule{1.405pt}{0.121pt}}
\multiput(1071.00,641.34)(119.084,81.000){2}{\rule{0.702pt}{0.800pt}}
\multiput(1193.00,725.41)(0.787,0.501){147}{\rule{1.457pt}{0.121pt}}
\multiput(1193.00,722.34)(117.976,77.000){2}{\rule{0.729pt}{0.800pt}}
\multiput(1314.00,802.41)(2.499,0.504){43}{\rule{4.104pt}{0.121pt}}
\multiput(1314.00,799.34)(113.482,25.000){2}{\rule{2.052pt}{0.800pt}}
\sbox{\plotpoint}{\rule[-0.200pt]{0.400pt}{0.400pt}}%
\put(220,520){\usebox{\plotpoint}}
\multiput(220,520)(20.756,0.000){59}{\usebox{\plotpoint}}
\put(1436,520){\usebox{\plotpoint}}
\put(220,673){\usebox{\plotpoint}}
\multiput(220,673)(20.756,0.000){59}{\usebox{\plotpoint}}
\put(1436,673){\usebox{\plotpoint}}
\end{picture}
    \caption{\label{frac}  }
\end {figure}

\clearpage
\begin{figure} [t]
\setlength{\unitlength}{0.240900pt}
\ifx\plotpoint\undefined\newsavebox{\plotpoint}\fi
\begin{picture}(1500,900)(0,0)
\font\gnuplot=cmr10 at 10pt
\gnuplot
\sbox{\plotpoint}{\rule[-0.200pt]{0.400pt}{0.400pt}}%
\put(220.0,113.0){\rule[-0.200pt]{0.400pt}{184.048pt}}
\put(220.0,113.0){\rule[-0.200pt]{4.818pt}{0.400pt}}
\put(198,113){\makebox(0,0)[r]{0.05}}
\put(1416.0,113.0){\rule[-0.200pt]{4.818pt}{0.400pt}}
\put(220.0,304.0){\rule[-0.200pt]{4.818pt}{0.400pt}}
\put(198,304){\makebox(0,0)[r]{0.1}}
\put(1416.0,304.0){\rule[-0.200pt]{4.818pt}{0.400pt}}
\put(220.0,495.0){\rule[-0.200pt]{4.818pt}{0.400pt}}
\put(198,495){\makebox(0,0)[r]{0.15}}
\put(1416.0,495.0){\rule[-0.200pt]{4.818pt}{0.400pt}}
\put(220.0,686.0){\rule[-0.200pt]{4.818pt}{0.400pt}}
\put(198,686){\makebox(0,0)[r]{0.2}}
\put(1416.0,686.0){\rule[-0.200pt]{4.818pt}{0.400pt}}
\put(220.0,877.0){\rule[-0.200pt]{4.818pt}{0.400pt}}
\put(198,877){\makebox(0,0)[r]{0.25}}
\put(1416.0,877.0){\rule[-0.200pt]{4.818pt}{0.400pt}}
\put(220.0,113.0){\rule[-0.200pt]{0.400pt}{4.818pt}}
\put(220,68){\makebox(0,0){0}}
\put(220.0,857.0){\rule[-0.200pt]{0.400pt}{4.818pt}}
\put(463.0,113.0){\rule[-0.200pt]{0.400pt}{4.818pt}}
\put(463,68){\makebox(0,0){2}}
\put(463.0,857.0){\rule[-0.200pt]{0.400pt}{4.818pt}}
\put(706.0,113.0){\rule[-0.200pt]{0.400pt}{4.818pt}}
\put(706,68){\makebox(0,0){4}}
\put(706.0,857.0){\rule[-0.200pt]{0.400pt}{4.818pt}}
\put(950.0,113.0){\rule[-0.200pt]{0.400pt}{4.818pt}}
\put(950,68){\makebox(0,0){6}}
\put(950.0,857.0){\rule[-0.200pt]{0.400pt}{4.818pt}}
\put(1193.0,113.0){\rule[-0.200pt]{0.400pt}{4.818pt}}
\put(1193,68){\makebox(0,0){8}}
\put(1193.0,857.0){\rule[-0.200pt]{0.400pt}{4.818pt}}
\put(1436.0,113.0){\rule[-0.200pt]{0.400pt}{4.818pt}}
\put(1436,68){\makebox(0,0){10}}
\put(1436.0,857.0){\rule[-0.200pt]{0.400pt}{4.818pt}}
\put(220.0,113.0){\rule[-0.200pt]{292.934pt}{0.400pt}}
\put(1436.0,113.0){\rule[-0.200pt]{0.400pt}{184.048pt}}
\put(220.0,877.0){\rule[-0.200pt]{292.934pt}{0.400pt}}
\put(45,495){\makebox(0,0){${\cal Z}$ }}
\put(828,23){\makebox(0,0){$\tau_{\mbox{ref}}$}}
\put(1314,449){\makebox(0,0)[l]{1 \%}}
\put(1314,564){\makebox(0,0)[l]{0.1 \%}}
\put(220.0,113.0){\rule[-0.200pt]{0.400pt}{184.048pt}}
\sbox{\plotpoint}{\rule[-0.400pt]{0.800pt}{0.800pt}}%
\put(220,247){\usebox{\plotpoint}}
\multiput(220.00,245.09)(3.328,-0.506){31}{\rule{5.337pt}{0.122pt}}
\multiput(220.00,245.34)(110.923,-19.000){2}{\rule{2.668pt}{0.800pt}}
\multiput(342.00,229.41)(0.798,0.501){145}{\rule{1.474pt}{0.121pt}}
\multiput(342.00,226.34)(117.941,76.000){2}{\rule{0.737pt}{0.800pt}}
\multiput(463.00,305.41)(1.625,0.503){69}{\rule{2.768pt}{0.121pt}}
\multiput(463.00,302.34)(116.254,38.000){2}{\rule{1.384pt}{0.800pt}}
\multiput(586.41,342.00)(0.501,0.553){235}{\rule{0.121pt}{1.086pt}}
\multiput(583.34,342.00)(121.000,131.746){2}{\rule{0.800pt}{0.543pt}}
\multiput(706.00,474.09)(3.328,-0.506){31}{\rule{5.337pt}{0.122pt}}
\multiput(706.00,474.34)(110.923,-19.000){2}{\rule{2.668pt}{0.800pt}}
\multiput(828.00,458.41)(1.076,0.502){107}{\rule{1.912pt}{0.121pt}}
\multiput(828.00,455.34)(118.031,57.000){2}{\rule{0.956pt}{0.800pt}}
\multiput(950.00,515.41)(1.612,0.503){69}{\rule{2.747pt}{0.121pt}}
\multiput(950.00,512.34)(115.298,38.000){2}{\rule{1.374pt}{0.800pt}}
\multiput(1071.00,553.41)(0.794,0.501){147}{\rule{1.468pt}{0.121pt}}
\multiput(1071.00,550.34)(118.954,77.000){2}{\rule{0.734pt}{0.800pt}}
\multiput(1193.00,630.41)(3.301,0.506){31}{\rule{5.295pt}{0.122pt}}
\multiput(1193.00,627.34)(110.011,19.000){2}{\rule{2.647pt}{0.800pt}}
\multiput(1314.00,649.41)(0.804,0.501){145}{\rule{1.484pt}{0.121pt}}
\multiput(1314.00,646.34)(118.919,76.000){2}{\rule{0.742pt}{0.800pt}}
\sbox{\plotpoint}{\rule[-0.200pt]{0.400pt}{0.400pt}}%
\put(220,419){\usebox{\plotpoint}}
\multiput(220,419)(20.756,0.000){59}{\usebox{\plotpoint}}
\put(1436,419){\usebox{\plotpoint}}
\put(220,533){\usebox{\plotpoint}}
\multiput(220,533)(20.756,0.000){59}{\usebox{\plotpoint}}
\put(1436,533){\usebox{\plotpoint}}
\end{picture}
    \caption{\label{refrac}  }
\end {figure}

\clearpage
\begin{figure} [t]
\setlength{\unitlength}{0.240900pt}
\ifx\plotpoint\undefined\newsavebox{\plotpoint}\fi
\begin{picture}(1500,900)(0,0)
\font\gnuplot=cmr10 at 10pt
\gnuplot
\sbox{\plotpoint}{\rule[-0.200pt]{0.400pt}{0.400pt}}%
\put(220.0,113.0){\rule[-0.200pt]{292.934pt}{0.400pt}}
\put(220.0,113.0){\rule[-0.200pt]{4.818pt}{0.400pt}}
\put(198,113){\makebox(0,0)[r]{0}}
\put(1416.0,113.0){\rule[-0.200pt]{4.818pt}{0.400pt}}
\put(220.0,266.0){\rule[-0.200pt]{4.818pt}{0.400pt}}
\put(198,266){\makebox(0,0)[r]{0.05}}
\put(1416.0,266.0){\rule[-0.200pt]{4.818pt}{0.400pt}}
\put(220.0,419.0){\rule[-0.200pt]{4.818pt}{0.400pt}}
\put(198,419){\makebox(0,0)[r]{0.1}}
\put(1416.0,419.0){\rule[-0.200pt]{4.818pt}{0.400pt}}
\put(220.0,571.0){\rule[-0.200pt]{4.818pt}{0.400pt}}
\put(198,571){\makebox(0,0)[r]{0.15}}
\put(1416.0,571.0){\rule[-0.200pt]{4.818pt}{0.400pt}}
\put(220.0,724.0){\rule[-0.200pt]{4.818pt}{0.400pt}}
\put(198,724){\makebox(0,0)[r]{0.2}}
\put(1416.0,724.0){\rule[-0.200pt]{4.818pt}{0.400pt}}
\put(220.0,877.0){\rule[-0.200pt]{4.818pt}{0.400pt}}
\put(198,877){\makebox(0,0)[r]{0.25}}
\put(1416.0,877.0){\rule[-0.200pt]{4.818pt}{0.400pt}}
\put(220.0,113.0){\rule[-0.200pt]{0.400pt}{4.818pt}}
\put(220,68){\makebox(0,0){2}}
\put(220.0,857.0){\rule[-0.200pt]{0.400pt}{4.818pt}}
\put(463.0,113.0){\rule[-0.200pt]{0.400pt}{4.818pt}}
\put(463,68){\makebox(0,0){3}}
\put(463.0,857.0){\rule[-0.200pt]{0.400pt}{4.818pt}}
\put(706.0,113.0){\rule[-0.200pt]{0.400pt}{4.818pt}}
\put(706,68){\makebox(0,0){4}}
\put(706.0,857.0){\rule[-0.200pt]{0.400pt}{4.818pt}}
\put(950.0,113.0){\rule[-0.200pt]{0.400pt}{4.818pt}}
\put(950,68){\makebox(0,0){5}}
\put(950.0,857.0){\rule[-0.200pt]{0.400pt}{4.818pt}}
\put(1193.0,113.0){\rule[-0.200pt]{0.400pt}{4.818pt}}
\put(1193,68){\makebox(0,0){6}}
\put(1193.0,857.0){\rule[-0.200pt]{0.400pt}{4.818pt}}
\put(1436.0,113.0){\rule[-0.200pt]{0.400pt}{4.818pt}}
\put(1436,68){\makebox(0,0){7}}
\put(1436.0,857.0){\rule[-0.200pt]{0.400pt}{4.818pt}}
\put(220.0,113.0){\rule[-0.200pt]{292.934pt}{0.400pt}}
\put(1436.0,113.0){\rule[-0.200pt]{0.400pt}{184.048pt}}
\put(220.0,877.0){\rule[-0.200pt]{292.934pt}{0.400pt}}
\put(45,495){\makebox(0,0){${\cal Z}$ }}
\put(828,23){\makebox(0,0){$\hat{s}$}}
\put(1314,541){\makebox(0,0)[l]{1 \%}}
\put(1314,633){\makebox(0,0)[l]{0.1 \%}}
\put(220.0,113.0){\rule[-0.200pt]{0.400pt}{184.048pt}}
\sbox{\plotpoint}{\rule[-0.400pt]{0.800pt}{0.800pt}}%
\put(220,846){\usebox{\plotpoint}}
\multiput(221.41,837.37)(0.500,-1.176){479}{\rule{0.121pt}{2.080pt}}
\multiput(218.34,841.68)(243.000,-566.683){2}{\rule{0.800pt}{1.040pt}}
\multiput(463.00,276.39)(26.919,0.536){5}{\rule{32.600pt}{0.129pt}}
\multiput(463.00,273.34)(175.337,6.000){2}{\rule{16.300pt}{0.800pt}}
\multiput(706.00,282.41)(1.615,0.501){145}{\rule{2.768pt}{0.121pt}}
\multiput(706.00,279.34)(238.254,76.000){2}{\rule{1.384pt}{0.800pt}}
\multiput(950.00,355.09)(0.998,-0.501){237}{\rule{1.793pt}{0.121pt}}
\multiput(950.00,355.34)(239.278,-122.000){2}{\rule{0.897pt}{0.800pt}}
\multiput(1193.00,236.41)(0.620,0.500){385}{\rule{1.192pt}{0.121pt}}
\multiput(1193.00,233.34)(240.526,196.000){2}{\rule{0.596pt}{0.800pt}}
\sbox{\plotpoint}{\rule[-0.200pt]{0.400pt}{0.400pt}}%
\put(220,510){\usebox{\plotpoint}}
\multiput(220,510)(20.756,0.000){59}{\usebox{\plotpoint}}
\put(1436,510){\usebox{\plotpoint}}
\put(220,602){\usebox{\plotpoint}}
\multiput(220,602)(20.756,0.000){59}{\usebox{\plotpoint}}
\put(1436,602){\usebox{\plotpoint}}
\end{picture}
    \caption{\label{states} }
\end {figure}

\end{document}